\documentclass[a4paper,fleqn,usenatbib]{mnras}

\usepackage[T1]{fontenc}
\usepackage{ae,aecompl}

\usepackage{graphicx}	% Including figure files
\usepackage{amsmath}	% Advanced maths commands
\usepackage{amssymb}	% Extra maths symbols

\usepackage{newtxtext,newtxmath}

\newcommand{\angstrom}{\mbox{\normalfont\AA}}

\title[AT2019pev]{An X-ray View of the Ambiguous Nuclear Transient AT2019pev}

\author[Yu et al.]{
Zhefu Yu$^{1}$\thanks{Email: yu.2231@osu.edu},
C.~S.~Kochanek$^{1,2}$,
S.~Mathur$^{1,2}$,
K.~Auchettl$^{3,4,5}$,
D.~Grupe$^{6}$,
T.~W.-S.~Holoien$^{7}$\thanks{NHFP Einstein Fellow}
\\
$^{1}$Department of Astronomy, The Ohio State University, Columbus, Ohio 43210, USA\\
$^{2}$Center of Cosmology and Astro-Particle Physics, The Ohio State University, Columbus, Ohio, 43210, USA\\
$^{3}$School of Physics, The University of Melbourne, Parkville, VIC 3010, Australia\\
$^{4}$ARC Centre of Excellence for All Sky Astrophysics in 3 Dimensions (ASTRO 3D)\\
$^{5}$Department of Astronomy and Astrophysics, University of California, Santa Cruz, CA 95064, USA\\
$^{6}$Department of Earth and Space Sciences, Morehead State University, Morehead, KY 40514, USA\\
$^{7}$The Observatories of the Carnegie Institution for Science, 813 Santa Barbara St., Pasadena, CA 91101, USA
}

\date{Accepted XXX. Received YYY; in original form ZZZ}

\pubyear{2021}

\begin{document}
\label{firstpage}
\pagerange{\pageref{firstpage}--\pageref{lastpage}}
\maketitle

\begin{abstract}
AT2019pev is a nuclear transient in a narrow-line Seyfert 1 galaxy at $z=0.096$. The archival ultraviolet, optical and infrared data showed features of both tidal disruption events (TDEs) and active galactic nuclei (AGNs), and its nature is not fully understood. We present detailed X-ray observations of AT2019pev taken with \textit{Swift}, Chandra and NICER over 173 days of its evolution since the first \textit{Swift} XRT epoch. The X-ray luminosity increases by a factor of five in five days from the first \textit{Swift} XRT epoch to the lightcurve peak. The lightcurve decays by a factor of ten over $\sim$75 days and then flattens with a weak re-brightening trend at late times. The X-ray spectra show a "harder-when-brighter" trend before peak and a "harder-when-fainter" trend after peak, which may indicate a transition of accretion states. The archival ground-based optical observations show similar time evolution as the X-ray lightcurves. Beyond the seasonal limit of the ground-based observations, the Gaia lightcurve is rising toward an equally bright or brighter peak 223 days after the optical discovery. Combining our X-ray analysis and archival multi-wavelength data, AT2019pev more closely resembles an AGN transient. 
\end{abstract}


%%%%%%%%%%%%%%%%%%%%%%%%%%%%%%%%%%%%%%%%%%%%%%%%%%

%%%%%%%%%%%%%%%%% BODY OF PAPER %%%%%%%%%%%%%%%%%%

%Introduction 
\section{Introduction} \label{sec:intro}
Supermassive black holes (SMBHs) ubiquitously exist at the centers of massive galaxies \citep[e.g.,][]{Richstone1998,Kormendy2013}. If it is accreting, the resulting active galactic nucleus (AGN) shows stochastic ultraviolet (UV)/optical continuum variability with a typical amplitude of $\sim 10\% - 20\%$ on the time scales of months \citep[e.g.,][]{Perola1982,Giveon1999,Peterson2001,Kelly2009,MacLeod2010}. A small fraction of AGNs exhibit more drastic changes that differ significantly from the normal stochastic variability, such as changing-look (CL) AGNs \citep[e.g.,][]{Bianchi2005, Shappee2014,2016MNRAS.457..389M,2022MNRAS.511...54H} and rapid turn-on events \citep[e.g.,][]{Gezari2017_turnonAGN,Frederick2019}. Transients associated with SMBHs also occur in quiescent galaxies. In particular, when the SMBH tidally disrupts a passing star, about half of the debris can be accreted to produce a luminous flare in a tidal disruption event \citep[TDE, e.g.,][]{Rees1988,Evans1989}. 

Both TDEs and AGN-associated transients provide a unique opportunity to study the accretion physics around SMBHs. While both are luminous flares, they have different spectral energy distributions (SEDs), photometric evolution and spectral properties. For example, the SEDs of TDEs are usually well described by a black-body with a temperature of a few $10^4$ K \citep[e.g.,][]{Holoien2019_PK18kh,Hinkle2021_swift}, while AGN SEDs are better described by a power-law \citep[e.g.,][]{VandenBerk2001}. The UV/optical lightcurves of AGNs show stochastic variability, while TDE lightcurves generally decay monotonically after peak without significant additional variability \citep[e.g.,][]{Holoien2014,Brown2017,Auchettl2018}, although there are TDEs that show re-brightening episodes \citep[e.g.,][]{Holoien2019_PK18kh,Wevers2019,Wevers2021,Hinkle2021,Payne2021}. The UV/optical spectra of TDEs have very broad H and/or He lines with ${\rm FWHM} \gtrsim 10^4$ km s$^{-1}$ \citep[e.g.,][]{Arcavi2014,Charalampopoulos2022}. TDEs can also show Bowen fluorescence features such as N III \citep[e.g.,][]{Leloudas2019,vanVelzen2021}. In contrast, the spectra of type I AGNs are characterized by broad Balmer lines with ${\rm FWHM} \sim 10^3$ km s$^{-1}$ and narrow forbidden lines \citep[e.g.,][]{VandenBerk2001}. 

X-ray observations are also critical in distinguishing TDEs and AGNs. Soft X-ray emission is predicted to contribute a large fraction of TDE luminosity \citep[e.g.,][]{Ulmer1999} and is observed in TDEs \citep[e.g.,][]{Auchettl2017,Brown2017,Gezari2017,Hinkle2021}. X-ray emission is also common for AGNs \citep[e.g.,][]{Haardt1991}. Similar to the UV and optical bands, the X-ray lightcurves of TDEs generally decay coherently after peak, again in contrast to the stochastic variability of AGNs \citep[e.g.,][]{Auchettl2017,Auchettl2018}. TDEs are usually less absorbed and have intrinsically softer spectra than AGNs \citep[e.g.,][]{Auchettl2018}. The X-ray hardness ratio of TDEs is roughly constant as the transient evolves, while AGNs generally show a "harder-when-fainter" behavior \citep[e.g.,][]{Shemmer2008,Auchettl2018}. 

Recent sky surveys, such as the All-Sky Automated Survey for Supernovae \citep[ASAS-SN, ][]{Shappee2014,Kochanek2017}, the Panoramic Survey Telescope and Rapid Response System \citep[Pan-STARRS, ][]{Chambers2016}, Gaia \citep{GaiaCollaboration2016}, the Asteroid Terrestrial Impact Last Alert System \citep[ATLAS, ][]{Tonry2018} and the Zwicky Transient Facility \citep[ZTF, ][]{Bellm2019}, are rapidly increasing the discovery of nuclear transients. These surveys have also discovered ambiguous nuclear transients (ANTs) that can be classified as neither TDEs nor AGNs. For example, \citet{Trakhtenbrot2019} proposed a new class of nuclear transients represented by AT2017bgt with a strong UV flare, Bowen fluorescence features and a slow UV decline of $\lesssim 0.7$ mag over 14 months. ASASSN-18jd \citep{Neustadt2020} exhibits both TDE-like features, such as an SED well fit by a black body model with $T \approx 2.5\times10^4$ K, and AGN-like features, such as the strong C III] $\lambda1909$ line. ASASSN-20hx \citep{Hinkle2021_20hx} showed a TDE-like SED that is well-fit by a $T \approx 21000$ K black-body, while it also showed AGN-like features such as an archival X-ray detection of the host-galaxy and an X-ray spectrum described by a power-law of index $\Gamma \sim 2.3$. However, the featureless UV/optical spectra of ASASSN-20hx were unusual for either scenario. 

The nuclear transient AT2019pev/ZTF19abvgxrq/Gaia19eby at $(\alpha, \delta)$ = $(04{\rm :}29{\rm :}22.72, 00{\rm :}37{\rm :}07.6)$ was first reported on 2019-09-01 (MJD 58727) on the Transient Name Server (TNS) by ZTF \citep{Forster2019TNS}. ATLAS, Gaia and Pan-STARRS also reported detection on 2019-09-04 (MJD 58730), 2019-09-13 (MJD 58739) and 2019-09-26 (MJD 58752). Early X-ray observations were reported by \citet{Gezari2019}, \citet{Kara2019}, \citet{Miller2019ATel}, \citet{Ferrigno2019ATel}, \citet{Mathur2019} and \citet{Chung2019ATel}. \citet{Frederick2021} presented UV, optical and infrared observations of AT2019pev along with optical spectra. Based on the width of the broad Balmer lines and the [O III]$/$H$\beta$ line ratios, they classified the host galaxy of AT2019pev as a narrow-line Seyfert 1 (NLSy1) galaxy at $z = 0.096$. AT2019pev showed features of both TDEs and AGNs. 

Although \citet{Frederick2021} classified it as an AGN-associated transient, they did not provide a detailed analysis of the available \textit{Swift} X-ray telescope (XRT) data or other available X-ray data. Here we present extensive X-ray observations by \textit{Swift}, Chandra and the Neutron Star Interior Composition Explorer (NICER) over 173 days from the first \textit{Swift} XRT epoch to further probe the nature of AT2019pev. We also report that AT2019pev seems to have had a second optical peak detected by Gaia, but full characterisation of this additional peak was missed as observations were Sun constrained. We discuss the X-ray observations in Section \ref{sec:obs}. Section \ref{sec:anl} describes our data analysis procedure. We present and discuss the results in the context of different scenarios in Section \ref{sec:discussion}. We summarize the paper and draw conclusions in Section \ref{sec:summary}. We have adopted a $\Lambda$CDM cosmology with $H_0 = 70 \, {\rm km/s/Mpc}$, $\Omega_m = 0.3$ and $\Omega_{\Lambda} = 0.7$ throughout the paper.

%Observations 
\section{Observations} \label{sec:obs}
We obtained X-ray observations from the \textit{Swift} XRT \citep[][]{Burrows2005}, Chandra High Resolution Camera (HRC)/Low Energy Transmission Grating \citep[LETG, ][]{Brinkman2000,Weisskopf2002}, Chandra Advanced Charged Coupled Device Imaging Spectrometer \citep[ACIS, ][]{Garmire1999} and NICER \citep{2012SPIE.8443E..13G}. We discuss the details of the observations from each instrument below.

%Obs: Swift
\subsection{\textit{Swift}} \label{subsec:obs_swift}
AT2019pev was observed using the \textit{Swift} XRT in photon counting mode (pc-mode, \citealt{Hill2004}) from 2019-09-24 (MJD 58750) to 2020-02-22 (MJD 58901). We obtained 45 epochs (Observation IDs: 00011566001, 00011566004-005, 00011566008-009, 00012028001-046) that had an exposure time ranging between 700s to 10\,ks giving a total exposure time of 104\,ks. Using the \textit{Swift} analysis tool \textsc{xrtpipeline}, all observations were reprocessed from level one XRT data using standard filter and screening criteria\footnote{\url{http://swift.gsfc.nasa.gov/analysis/xrt_swguide_v1_2.pdf}} to produce cleaned event files and exposure maps.

We obtained a background subtracted count rate for each observation using a source region with a radius of 50'' centered on the position of AT2019pev and a 150'' radius source free background region centered at $(\alpha,\delta)$ = $(04{\rm :}29{\rm :}22.05, +00{\rm :}37{\rm :}47.51)$. All count rates were corrected for encircled energy fraction. To increase the signal to noise of our observations, we
combined our individual Swift epochs into 12 time bins using \textsc{XSELECT} version 2.4g, allowing us to extract spectra with $>1000$ background subtracted counts.

Both source and background spectra were extracted from these merged observations using \textsc{xrtproducts} version 0.4.2 and the regions defined above. We extracted ancillary response files (ARF) using the task \textsc{xrtmkarf} after we merged individual exposure maps that were generated by the \textit{xrtpipeline} using the FTOOLS program \textsc{XIMAGE}. We used the ready-made response matrix files (RMFs) that are available with the \textit{Swift} calibration files. Each spectrum was grouped to have a minimum of 5 counts per energy bin using the FTOOLS command \textsc{grppha}.

%Obs: Chandra
\subsection{Chandra} \label{subsec:obs_chandra}
We obtained four epochs of Chandra observations (PI: Kochanek) on 2019-10-14 (MJD 58770, observation IDs 21404, 22875), 2019-10-27 (MJD 58783, observation IDs 21405, 22906), 2019-11-11 (MJD 58798, observation ID 21406) and 2019-11-22 (MJD 58809, observation IDs 21407, 23077, 23078). The observations were obtained with the low-energy transmission grating (LETG) with the high-resolution camera  for spectroscopy (HRC-S) as the detector. The LETG/HRC-S exposure times were 93 ks, 99 ks, 98 ks and 99 ks, respectively. 
We reprocessed the Chandra data  with Chandra Interactive Analysis of Observations (CIAO, \citealt{fru06}) package and extracted the grating spectra following the standard procedure in CIAO v4.10\footnote{\url{https://cxc.harvard.edu/ciao/}}. The spectral orders are overlapped on the same  dispersion axis, and the energy resolution of the detector can, in principle, be used to separate the orders. However, HRC-S does not have good enough spectral resolution for clear order sorting; therefore we included grating orders 1--8 in our analysis and combined the plus and minus orders. The results from the first epoch were reported in \citet{Mathur2019}. We binned the Chandra grating spectra with 100 channels in each bin.

We also obtained three additional epochs using the Chandra ACIS in VFAINT mode on 2019-10-12 (MJD 58768), 2019-11-12 (MJD 58799) and 2020-01-18 (MJD 58866) under the observation IDs 21390, 21391, and 21392 (PI: Auchettl).  The observations have exposure times of 30 ks, 28 ks and 30 ks, respectively. Standard data reduction and cleaning were performed using the CIAO command \textsc{chandra\_repro}. We also extracted spectra from each cleaned and reprocessed observation using the CIAO command \textsc{specextract}. All spectra were grouped using a minimum of 20 counts per bin.

%Obs: NICER
\subsection{NICER} \label{subsec:obs_nicer}
AT2019pev was also observed using NICER. In total, 41 epochs were taken between 2019-09-25 (MJD 58751) to 2020-03-15 (MJD 58923), totaling 66 ks of cumulative exposure (Observation IDs:2200860101-2200860143, 3200860101-3200860105, PI: Gendreau). The data were reduced using the NICERDAS version 6a. We applied standard filtering criteria similar to those used by \citet{2019ApJ...887L..25B} and \citet{Hinkle2021}\footnote{see \url{https://heasarc.gsfc.nasa.gov/docs/nicer/data_analysis/nicer_analysis_guide.html} for more details about these criteria.} and produced cleaned event files using the NICERDAS task \textsc{NICERL2}. Time averaged spectra and counts rates were extracted using XSELECT and we used the ready made ARF (nixtiaveonaxis20170601v004.arf) and RMF (nixtiref20170601v002.rmf) files that are made available with the NICER CALDB. All spectra were grouped with a minimum of 20 counts per energy bin, while background spectra were generated using the background modelling tool \textsc{NIBACKGEN3c50}\footnote{\url{https://heasarc.gsfc.nasa.gov/docs/nicer/tools/nicer_bkg_est_tools.html}}.

As the signal-to-noise ratio (SNR) of the NICER spectra drops significantly in the later epochs due to the shorter exposure times and lower source luminosity, we co-added the NICER epochs using \textsf{addspec}\footnote{\url{https://heasarc.gsfc.nasa.gov/lheasoft/ftools/fhelp/addspec.html}} so that the exposure time is at least 2 ks for each combined epoch. This gave us a total of 22 epochs that we used for our analysis.

%Data Analysis 
\section{Data Analysis} \label{sec:anl}

We analyzed the X-ray spectra over the $0.3 - 5$ keV energy range using \textsf{xspec} version 12.12.1 \citep{xspec}. In the Chandra grating and NICER spectra, there are low-SNR channels at the energy boundaries of the spectra due to the drop in the effective area. We excluded \textit{all} channels beyond the first channel where the SNR drops below 1.5. Since the low-SNR channels tend to have low flux, this minimizes potential biases toward high flux from only excluding low-SNR channels. 

We adopted the interstellar medium (ISM) absorption model \textsf{tbabs} \citep{tbabs}. We were unable to well constrain the absorption column density N(H) for most \textit{Swift}, Chandra and NICER epochs except for 11 early, high-SNR NICER epochs. Using these observations, we find that the column densities were consistent with the Galactic value of $6.04 \times 10^{20} \, {\rm cm}^{-2}$ from \citet{Galactic_nH} within the uncertainties and showed no evidence for time evolution, similar to the lack of N(H) evolution seen in TDEs \citep[e.g.,][]{Auchettl2017}. We therefore fixed N(H) to the Galactic value for our final analyses.   

We fit all spectra using a power-law (\textsf{powerlaw}), power-law plus back-body (\textsf{bbody}) and a power-law plus accretion disk multi-temperature black-body (\textsf{diskpbb}) model with $\chi^2$ statistics\footnote{see \url{https://heasarc.gsfc.nasa.gov/docs/xanadu/xspec/index.html} for details of the models.}. The disk model gave similar $\chi^2$ fit statistics to the black-body model so we did not include it in our final analyses and discussion. For the power-law + black-body model, the black-body components cannot be well constrained in some low-SNR spectra. For any epoch where the uncertainty of the black-body temperature $kT$ was larger than 20\%, we applied a Gaussian prior to $kT$ centered on the median temperature of the other epochs with a dispersion equal to half the difference between their 16$^{\rm th}$ and 84$^{\rm th}$ percentiles. When calculating or applying the prior, we excluded the epochs before the X-ray lightcurve peak, as these epochs have a different spectral shape. We also excluded the two NICER epochs with potential background subtraction issues discussed below. 

We derived the X-ray luminosity over the 0.3$-$5 keV energy range based on the best-fit models. We calculated the hardness ratio ${\rm HR} = (H - S) / (H + S)$, where $S$ and $H$ are the integrated fluxes between 0.3$-$2 keV and 2$-$5 keV, respectively. We did not use the more conventional 2$-$10 keV hard band because the 5$-$10 keV flux would simply be an extrapolation of the model given our data. Table \ref{tab:param} summarizes the luminosities, model parameters and hardness ratios based on the power-law + black-body model for all epochs ordered by $t = {\rm MJD} - {\rm MJD}_0$, where ${\rm MJD}_0 = 58727.5$ is the discovery date. The X-ray peak occurs at $t = 27.6$ days.

\begin{figure*}
\includegraphics[width=0.95\linewidth]{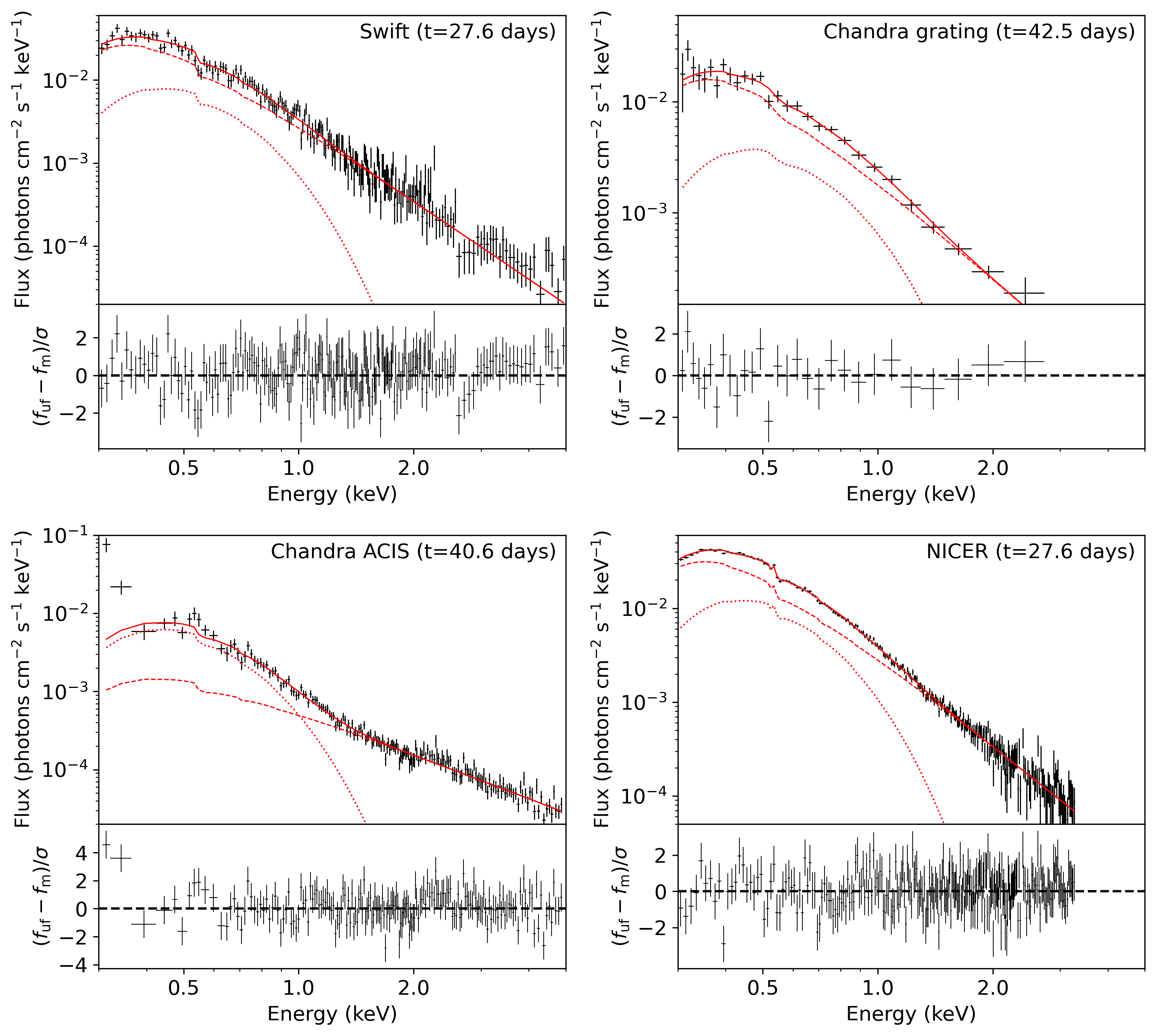}
\caption{Examples of the spectra and the power-law + black-body models for each instrument. The instrument name and the time since discovery are given in the upper right corner of each panel. The X-ray peak is at $t = 27.6$ days. The black crosses in the upper panels show the unfolded spectra. The solid red line shows the best-fit model, where the dash and dotted lines represent the power-law and black-body components, receptively. The lower panels show the normalised residual defined as the difference between the unfolded spectra $f_{\rm uf}$ and the best-fit model $f_{\rm m}$ divided by the uncertainty $\sigma$.}
\label{fig:fitting}
\end{figure*}

The observed X-ray spectrum is a ``folded'' version of the source spectrum due to the non-diagonal response matrix and the energy-dependent effective area of each instrument. We derived the unfolded spectra in \textsf{xspec} using the ratio between the folded and unfolded version of the best-fit model. Figure \ref{fig:fitting} shows examples of the unfolded spectra and the best-fit power-law + black-body models for each instrument. These aspects of the analysis apply to all instruments, and we next discuss details peculiar to each instrument. 

%Spectra fitting - Swift
\subsection{Swift} \label{subsec:anl_specfit_swift}
We can reasonably fit the \textit{Swift} spectra using either a single power-law with $\chi^2_r \sim 0.9 - 1.2$ or a power-law + black-body model with $\chi^2_r \sim 0.8 - 1.1$, where $\chi^2_r$ is the reduced $\chi^2$ statistic. For the latter model, the black-body temperature is well constrained for most epochs except for Sw08-02 (MJD 58763, $t=35.1$ days) and Sw21-29 (MJD 58832, $t=104.5$ days). We obtained a power-law index $\Gamma \sim 2.5 - 3.5$ for the single power-law model and found a $\Gamma \sim 2 - 3$ combined with a black-body temperature $kT \sim 0.12 - 0.15\, {\rm keV}$ for the power-law + black-body model. We used the F-test to compare the goodness of the fit of different models, and we defined a model to be significantly better if the F-test probability $p < 2 \times 10^{-4}$. The power-law + black-body model does not give a significantly better fit than the single power-law for any epochs except Sw00011566004 (MJD 58755, $t=27.6$ days). Figure \ref{fig:fitting} shows an example of the unfolded \textit{Swift} spectrum at $t=27.6$ days and the best-fit power-law + black-body model in the upper left panel. The model fits the spectra well with no systematic trend in the residuals. The power-law component dominates the high-energy spectra while both components contribute to the low-energy spectra. 

The epoch Sw00011566001 (MJD 58750, $t=22.7$ days), 5 days before the X-ray peak, has a much steeper spectrum than the other epochs with $\Gamma \sim 4.7$ ($\chi^2_r \sim 1.2$) for a power-law or $kT \sim 0.1\, {\rm keV}$ ($\chi^2_r \sim 1.1$) for a single black-body model. When fitting the black-body model, the spectrum does not show additional power-law components at high energies, and the power-law index is poorly constrained when fitting with the power-law + black-body model. The flux ratio of the power-law to the black-body then increases to $\sim 3 - 4$ at the lightcurve peak and is then roughly constant within the large uncertainties.

%Spectra fitting - Chandra
\subsection{Chandra} \label{subsec:anl_specfit_chandra}
We obtained a reasonable fit to these data with either a power-law ($\chi^2_r \sim 0.6 - 1.4$) or a power-law + black-body model ($\chi^2_r \sim 0.4 - 0.8$). We applied the Gaussian prior to the black-body temperature $kT$ for two epochs where $kT$ was poorly constrained. We obtained $\Gamma \sim 3$ for the single power-law model and $\Gamma \sim 1.6 - 3$ for the power-law + black-body model. The power-law + black-body model does not provide a significantly better fit than a single power-law. 

The Chandra ACIS spectra can only be well fit by the power-law + black-body model with $\chi^2_r \sim 0.8 - 1.2$, and the black-body temperature is well-constrained for all three epochs. Removing either component would lead to significantly worse fit with $\chi^2_r \sim 1.6 - 3.2$. We obtained $\Gamma \sim 2$ and $kT \sim 0.12 - 0.15\, {\rm keV}$. The upper right and lower left panels of Figure \ref{fig:fitting} show examples and the best-fit models for the Chandra grating spectra and ACIS spectra. 

%Spectra fitting - NICER
\subsection{NICER} \label{subsec:anl_specfit_nicer}
The NICER spectra can only be well fit by the power-law + black-body model with $\chi^2_r \sim 0.7 - 1.2$ except NI3200860104 (MJD 58923, $t=195.7$ days) where a single power-law also provides a reasonable fit with $\chi^2_r \sim 1$. We can well constrain the black-body temperature for all epochs except this one. We typically found $\Gamma \sim 2.5 - 3.5$ and $kT \sim 0.12 - 0.17\, {\rm keV}$. Figure \ref{fig:fitting} shows an example of a NICER spectrum and its model in the lower right panel. 

Two NICER epochs show uncommon spectral features at the energy boundaries that are likely due to problems in the estimated background. The merged spectrum of NI16-17 (Observation IDs 2200860119, 2200860120, MJD 58782, $t=54.3$ days) shows significant excess above 2 keV relative to the neighboring epochs that lead to a model with an extremely flat power-law component. The epoch NI3200860104 shows an excess at low energies. We excluded these two epochs from further analysis.

%Results and Discussion 
\section{Results and Discussion} \label{sec:discussion}
\begin{figure*}
\includegraphics[width=0.95\linewidth]{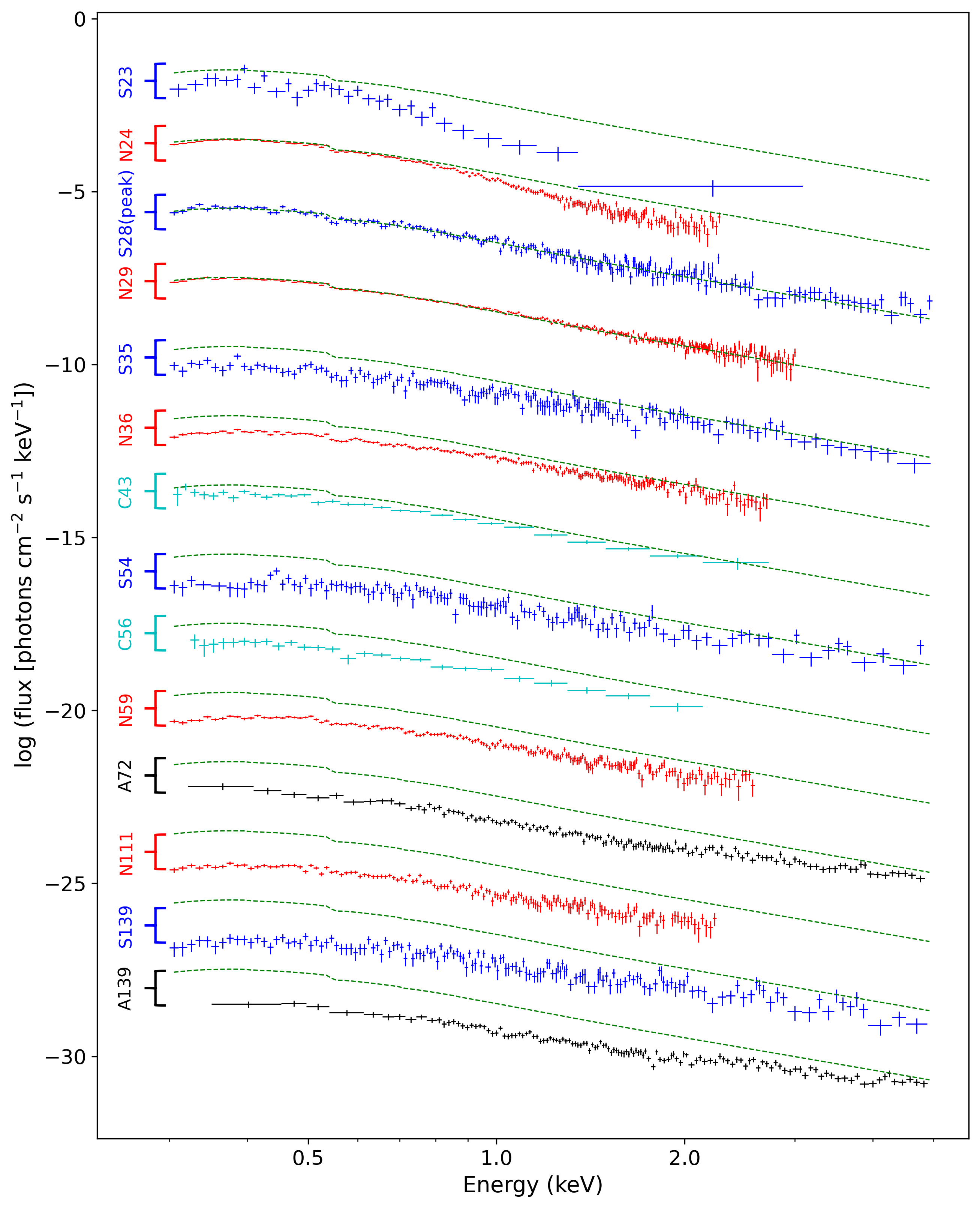}
\caption{Unfolded spectra for a sub-sample of the epochs. We applied constant shifts of 2 dex to the y-axis for visibility. The blue, cyan, black and red crosses marked by the letter ``S'', ``C'', ``A'' and ``N'' represent the spectra from \textit{Swift}, Chandra grating, Chandra ACIS and NICER, respectively. The X-ray spectrum corresponding to the peak (3rd from top) is labeled. The numbers after the letter markers represent the rounded time in days since discovery. The green dashed lines show the best-fit model for the epoch ``S28'' (Sw00011566004) at the X-ray lightcurve peak with a power-law index $\Gamma = 3.1$ and a black-body temperature $kT = 0.136$ keV to help illustrate the spectral changes.}
\label{fig:ufspec}
\end{figure*}

\begin{figure*}
\includegraphics[width=0.85\linewidth]{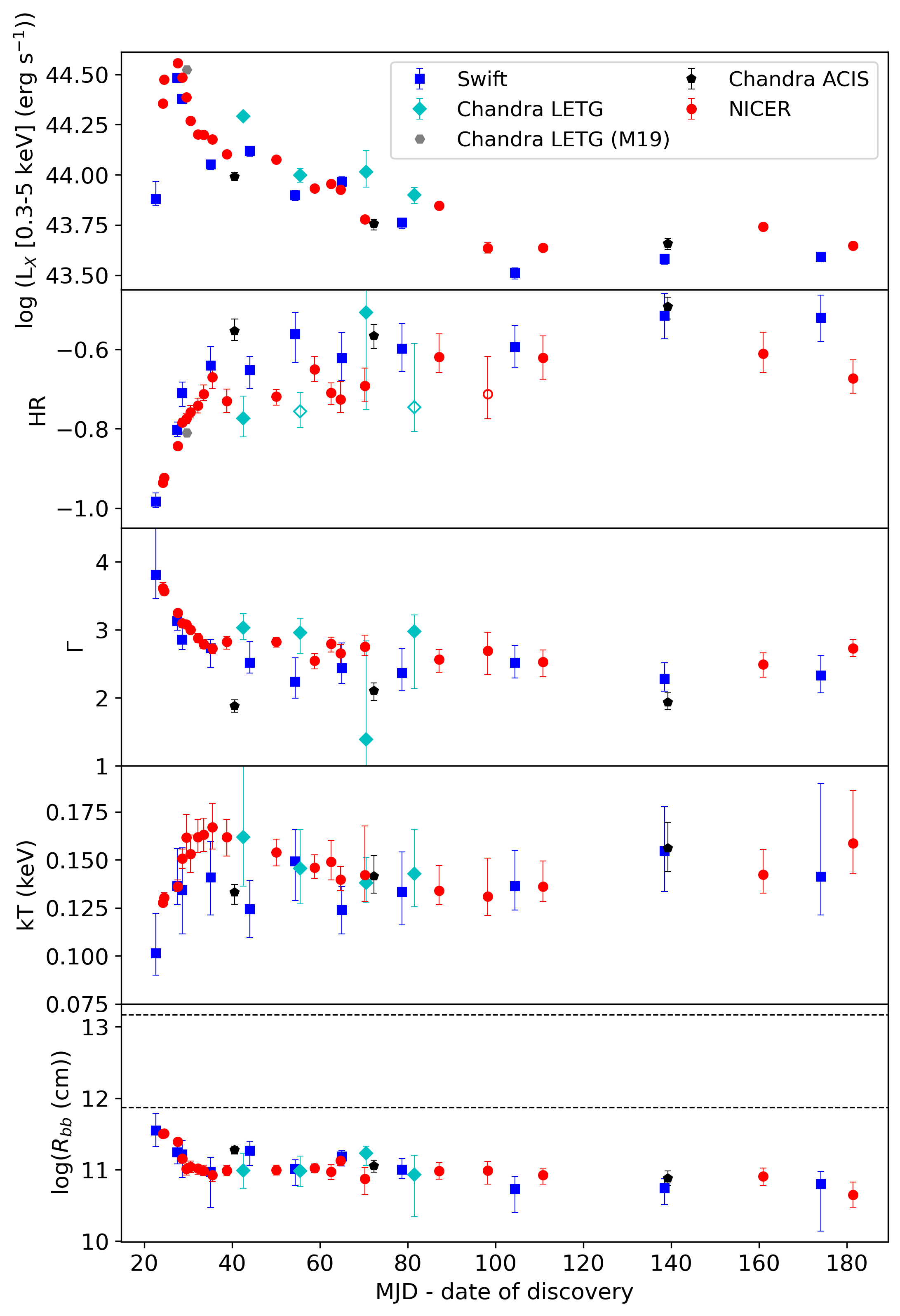}
\caption{The evolution of the 0.3$-$5 keV luminosity (top), the hardness ratio (upper middle), the power-law index $\Gamma$ (middle), the black-body temperature $kT$ (lower middle) and the black-body radius $R_{\rm bb}$ (bottom) with time. The blue squares, cyan diamonds, black pentagons and red circles represent the results from \textit{Swift}, Chandra LETG, Chandra ACIS and NICER spectra, respectively. The grey hexagons in the top two panels represent the Chandra LETG epoch from \citet{Miller2019ATel}. The dashed lines in the bottom panel are drawn at the Schwarzschild radii of the two black hole mass estimates $10^{7.7} M_\odot$ and $10^{6.4} M_\odot$ by \citet{Frederick2021}}
\label{fig:pvar}
\end{figure*}

\begin{figure}
\includegraphics[width=\linewidth]{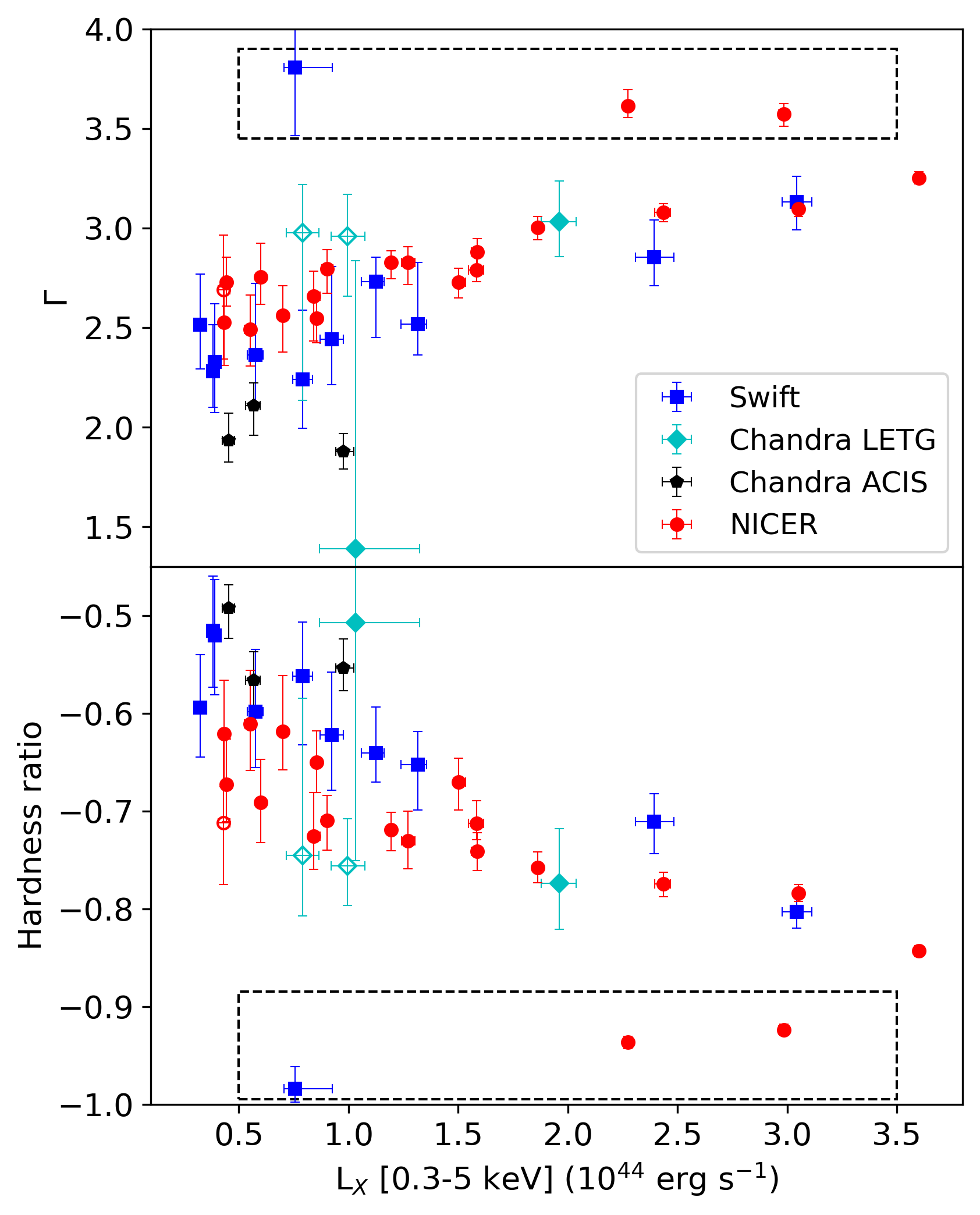}
\caption{The power-law index (upper panel) and the hardness ratio (lower panel) as a function of X-ray luminosity. The blue squares, cyan diamonds, black pentagons and red circles represent the results from \textit{Swift}, Chandra LETG, Chandra ACIS and NICER spectra, respectively. The three epochs obtained before the lightcurve peak are enclosed by the dashed box.}
\label{fig:gamma_HR_L}
\end{figure}

We adopt the power-law + black-body model as the fiducial model for our discussion, since it fits all epochs and is the only model that fits the Chandra ACIS and NICER spectra. Figure \ref{fig:ufspec} shows the unfolded spectra for a subset of epochs to illustrate how the spectra change over time. Figure \ref{fig:pvar} shows the evolution of the X-ray luminosity, the hardness ratio (HR), the power-law index $\Gamma$, the black-body temperature $kT$ and the black-body radius $R_{\rm bb}$ as a function of time. It also shows the luminosity and HR of an additional Chandra LETG epoch from \citet{Miller2019ATel}. The power-law component becomes harder with a flatter power-law index and a higher hardness ratio as the lightcurve decays. This can also be directly seen in Figure \ref{fig:ufspec}. Figure \ref{fig:gamma_HR_L} shows the power-law index $\Gamma$ and the HR as a function of the luminosity $L_X$. Most epochs show a positive (negative) correlation between $\Gamma$ (HR) and $L_X$, i.e., a "harder-when-fainter" behavior. The black-body temperature $kT$ shows a very weak increasing trend while the black-body radius $R_{\rm bb}$ declines over the early epochs. These trends flatten after $t \sim 40$ days.  

The host galaxy of AT2019pev was classified as a NLSy1 galaxy by \citet{Frederick2021}. However, NLSy1s may also host non-AGN transients, such as the TDE candidate PS16dtm \citep{Blanchard2017} and two other sources from \citet{Frederick2021} which they classified as TDEs. We next discuss the X-ray results along with archival multi-wavelength data in the context of both TDEs and AGNs. We then compare AT2019pev with other ANTs and discuss the features that are unusual for both scenarios and their implications. 

\subsection{The TDE Scenario} \label{subsec:discussion_tde}
\begin{figure*}
\includegraphics[width=0.89\linewidth]{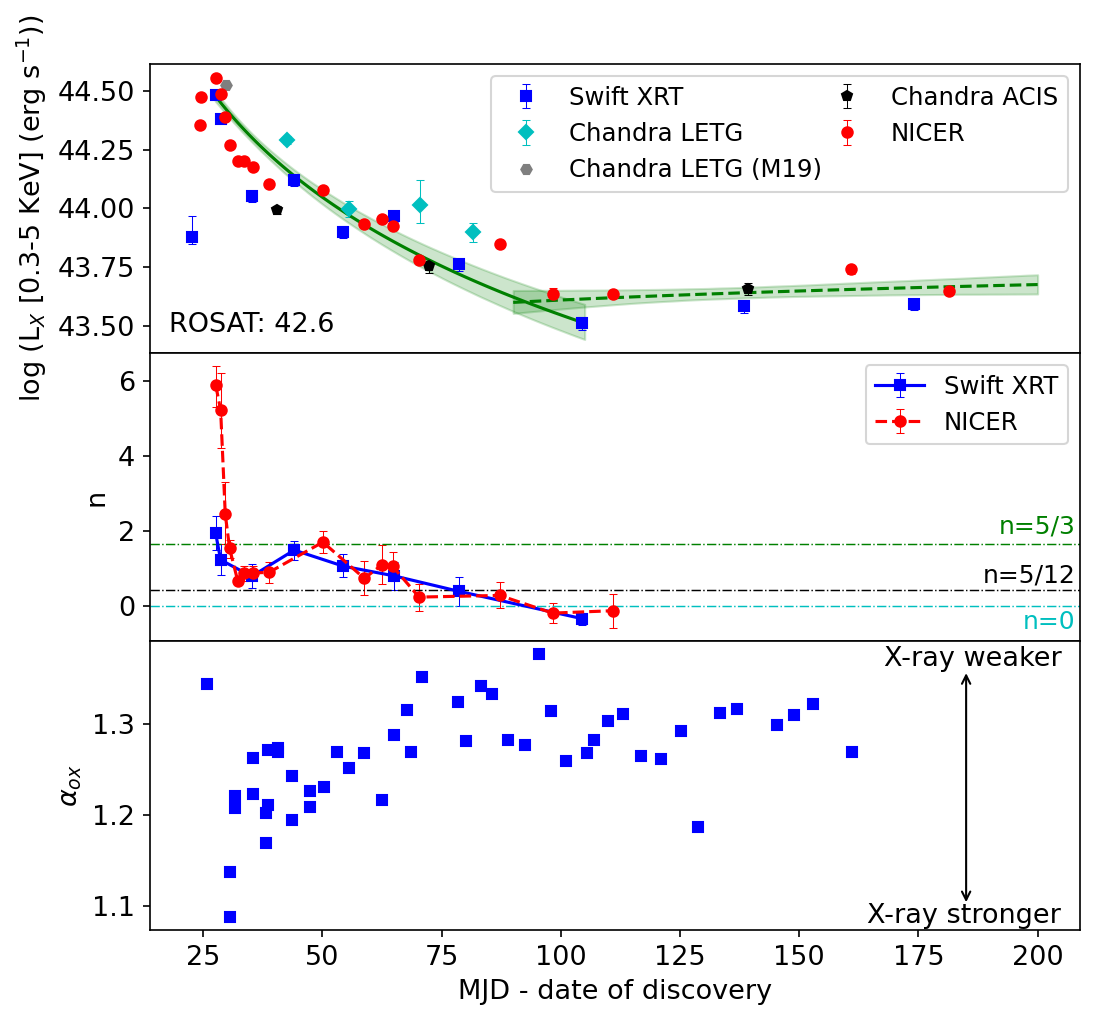}
\caption{({\it Top panel}) X-ray lightcurves. The symbols have the same meanings as Figure \ref{fig:pvar}. The green solid and dashed lines are the best-fit power-law models for the epochs between $t=27.5 - 105$ days and $t=90 - 200$ days, respectively, where the shaded regions are the 1$\sigma$ uncertainty bands. ({\it Middle panel}) Lightcurve power-law index $n$ as a function of time. The blue and red points are the results from \textit{Swift} and NICER lightcurves, respectively. The lines simply connect the data points to help illustrate the trend. The horizontal lines are drawn at $n = 0,\, 5/12,\, 5/3$ ({\it Bottom panel}) Optical-to-X-ray slope $\alpha_{\rm OX}$ by \citet{Frederick2021}. }
\label{fig:lcfit}
\end{figure*}

\begin{figure*}
\includegraphics[width=0.95\linewidth]{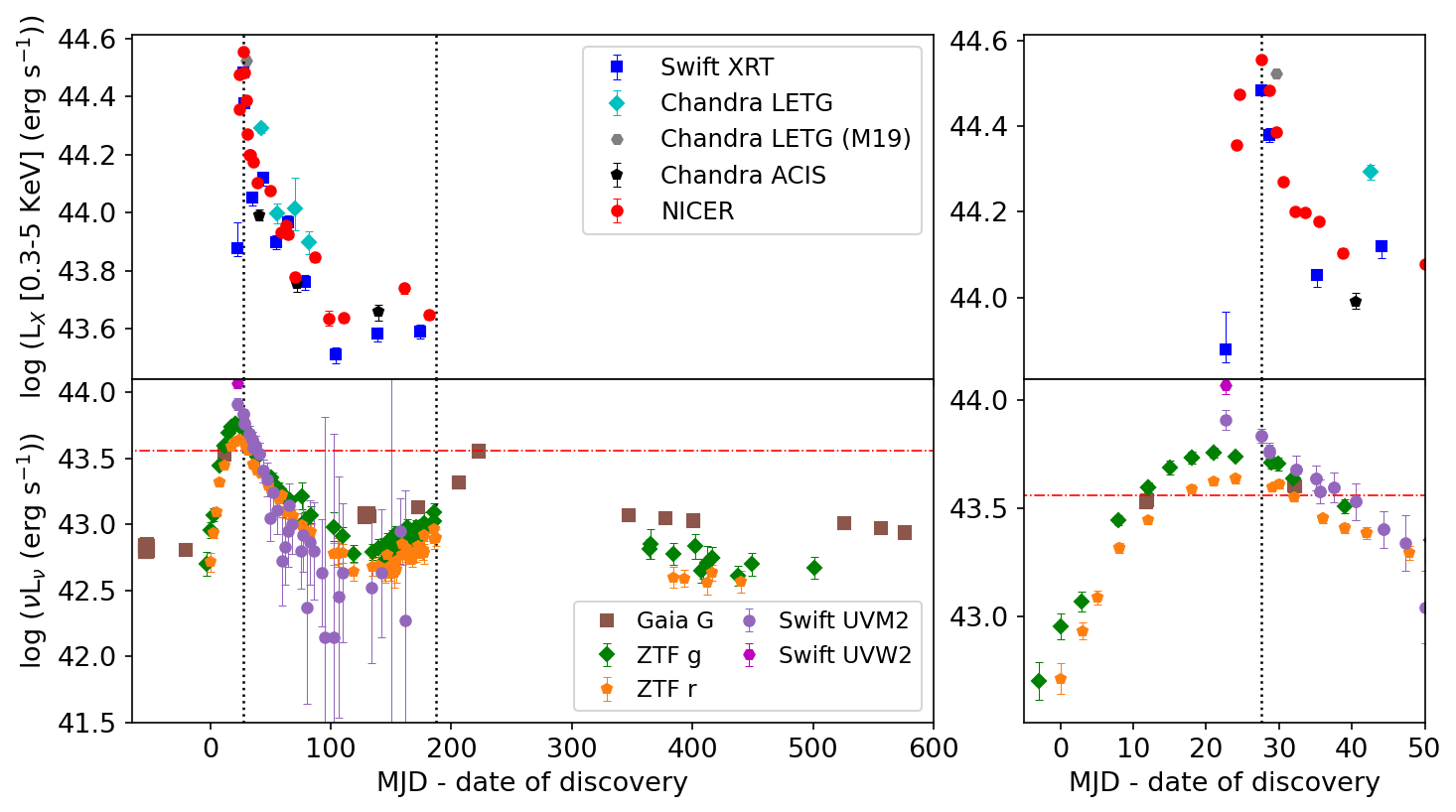}
\caption{({\it Upper row}) X-ray lightcurves. The symbols have the same meanings as Figure \ref{fig:pvar}. The right column is a zoom in around the X-ray peak with different y-axis scales from the left column. The black dotted lines are drawn at the peak of the X-ray lightcurve ($t=27.6$ days) and the last epoch of the ZTF lightcurves ($t=187.5$ days). ({\it Lower row}) UV and optical lightcurves. The brown, green, orange, purple and magenta points are the photometry in Gaia G, ZTF $g$, ZTF $r$, \textit{Swift} UVM2 and \textit{Swift} UVW2 bands, respectively. The Swift UVOT and the ZTF lightcurves are from \citet{Frederick2021}. The red dashed-dotted are drawn at the Eddington limit based on the virial black hole mass $M_{\rm BH, vir} = 10^{7.7} M_\odot$ by \citet{Frederick2021}, which can change by orders of magnitudes if adopting the black hole mass $M_{{\rm BH},M_r} = 10^{6.4} M_\odot$ based on the host galaxy luminosity.}
\label{fig:lcopt}
\end{figure*}

The top panels of Figures \ref{fig:pvar}, \ref{fig:lcfit} and \ref{fig:lcopt} show the X-ray lightcurve of AT2019pev. The first \textit{Swift} epoch ($t=22.7$ days) is about an order of magnitude brighter than the archival ROSAT detection \citep{Dickey1990,White1994a,White1994b,Gezari2019}. The luminosity then increases by a factor of 5 from the first \textit{Swift} epoch to the peak ($t=27.6$ days). It then drops by factors of 5 and 10 over the first 30 days and 75 days after the peak, respectively, and finally flattens after $t \sim 105$ days.  

Theoretical studies predict that the TDE lightcurves decay with a power-law $L \propto t^{-n}$, where the index $n$ depends on the properties of the star and the black hole as well as the details of the interaction \citep[e.g.,][]{Evans1989,Ramirez-Ruiz2009,Lodato2011,Coughlin2019,Gafton2019}. We fit the X-ray lightcurves with a power-law model. We inflated the log-space uncertainties of each epoch by a constant factor during the fitting until the reduced $\chi^2$ equals one, so that the model parameter uncertainties account for the scatter of the observed lightcurves. The green solid and dashed lines in the top panel of Figure \ref{fig:lcfit} show the best-fit power-law models for the lightcurves between $t=27.6 - 105$ days and between $t=90 - 200$ days, respectively. We obtained $n = 1.67 \pm 0.14$ for $t=27.6 - 105$ days, which is close to the conventional 5/3 predicted for TDEs based on the rate at which the gas returns to the pericenter when disrupting a main sequence star \citep[e.g.,][]{Evans1989}. We obtained $n = -0.22 \pm 0.23$ for $t=90 - 200$ days. This flat or weakly brightening lightcurve is inconsistent with all proposed decay models. For a solar-type star the fallback time is $t_{\rm fb} = 41 (M_{\rm BH}/10^6 M_\odot)^{1/2}$ days \citep[e.g.,][]{Gezari2017}. The virial and the host galaxy luminosity estimates of the black hole mass $M_{\rm BH, vir} = 10^{7.7} M_\odot$ and $M_{{\rm BH},M_r} = 10^{6.4} M_\odot$ by \citet{Frederick2021} yield $t_{\rm fb,vir} = 290$ days and $t_{{\rm fb},M_r} = 65$ days, respectively. This wide range of $t_{\rm fb}$ cannot constrain whether the lightcurve flattening after $t \sim 105$ days agrees with the scenario that the emission becomes disk dominated like that seen in TDEs. 

The X-ray lightcurve before $t \sim 105$ days shows an overall decaying trend with weak indications of a ``plateau'' near $t \sim 60 - 80$ days. To quantify the coherence of the lightcurve decay, we derived the power-law index $n$ as a function of time using a similar method to \citet{Auchettl2017,Auchettl2018}. For the $i^{\rm th}$ epoch ($i=1,2,3,...$) starting from the peak, we derived the power-law index $n$ by fitting the $i$ to $i+4$ epochs, stopping with a last measurement where there were only 3 epochs to fit. We fit each instrument separately to avoid artificial variability on short time scales due to the systematic difference in their absolute calibrations \citep[e.g.,][]{Madsen2017}. We did not fit the Chandra epochs given the small number of epochs. The middle panel of Figure \ref{fig:lcfit} shows the power-law index $n$ as a function of time. The lightcurve decays steeply near the peak and becomes flatter as it evolves. This quick drop of $n$ in the early epochs is inconsistent with TDEs, where $n$ varies little with time \citep{Auchettl2017,Auchettl2018}.

The bottom panel of Figure \ref{fig:lcfit} shows the optical-to-X-ray slope $\alpha_{\rm OX} = -0.3838 {\rm log}(L_{2\,{\rm keV}}/L_{2500\,\mbox{\scriptsize\AA}})$ \citep[e.g.,][]{Tananbaum1979} from \citet{Frederick2021}, where $L_{2\,{\rm keV}}$ and $L_{2500\,\mbox{\scriptsize\AA}}$ are the luminosities at 2 keV and 2500 \angstrom\,, respectively. The slope $\alpha_{\rm OX}$ varies between $\sim 1.1 - 1.3$, which is at the low end of the $\alpha_{\rm OX}$ distribution of the TDE sample from \citet{Wevers2020}. It is consistent with the hard state of the TDE candidate AT2018fyk \citep{Wevers2019,Wevers2021} but significantly harder than some other TDEs such as ASASSN-15oi with $\alpha_{\rm OX} \sim 2.0 - 2.5$ \citep[e.g.,][]{Holoien2018,Wevers2020}. The lower panel of Figure \ref{fig:lcopt} shows the ZTF and the \textit{Swift} Ultraviolet/Optical Telescope (UVOT, e.g., \citealt{Roming2005}) lightcurves from \citet{Frederick2021} and the Gaia G-band lightcurve from the Gaia Photoelectric Science Alerts \citep{GaiaCollaboration2016,Hodgkin2021}. To quantify the width the of the lightcurve peaks, we calculated the full-width half maximum (FWHM) as the difference between the two times where the linear interpolated luminosity equals half the peak luminosity. The UV/optical lightcurves have a wider peak with ${\rm FWHM} \approx 35$ days than the X-ray lightcurve with ${\rm FWHM} \approx 7$ days. The ZTF lightcurves peak at $t \sim 21 - 24$ days, which roughly coincides the first \textit{Swift} XRT epoch at $t=22.7$ days and slightly precedes the X-ray lightcurve peak at $t=27.6$ days (see the right column of Figure \ref{fig:lcopt}). 

The UV/optical lightcurves decline by about an order of magnitude after the peak until $t \sim 105$ days, similar to the X-ray lightcurve. The ZTF lightcurves then re-brighten by $\sim 0.5$ dex near the end of the observational season, in contrast to the flat or weakly re-brightening X-ray lightcurve. The Gaia lightcurve extends beyond the seasonal limit of the ground-based observations and is still rising and approaching the optical luminosity of the first peak just as the observations become Sun constrained. When next observable, it has again faded. The UV/optical re-brightening with a such strong secondary peak disfavors the TDE scenario, although there are exceptions. Some TDEs show plateau or re-brightening after the initial decay, such as PK18kh \citep{Holoien2019_PK18kh}, AT2018fyk \citep{Wevers2019,Wevers2021} and ASASSN-19dj \citep{Hinkle2021}. The TDE ASASSN-19bt shows a hump $\sim 30$ days before the major peak \citep{Holoien2019_19bt}. ASASSN-15lh, which has been classified as a super-luminous supernovae \citep[SLSNe-I, e.g.,][]{Dong2016,GodoyRivera2017} or a TDE \citep[e.g.,][]{Leloudas2016,Mummery2020}, also shows a re-brightening. The nuclear transient ASASSN-14ko shows periodic outbursts that are interpreted as repeated partial TDEs \citep{Payne2021}. The nuclear transient PS1-10adi re-brightens $\sim 1500$ days after the optical peak, which can be interpreted by the dust echo of a TDE in an AGN \citep{Kankare2017,Jiang2019}. The $g-r$ color of AT2019pev varies little with time, which is more consistent with the TDE scenario, as the AGN variability generally shows a "bluer-when-brighter" behavior \citep[e.g.,][]{Wilhite2005}.

The X-ray spectra for the early epochs have power-law indices of $\Gamma \sim 3 - 4$. This is consistent with TDEs, which tend to be quite soft events \citep[e.g.,][]{Auchettl2017,Auchettl2018}. However, the spectra harden as the transient fades. This is uncommon for TDEs where the hardness ratios vary little with time \citep{Auchettl2018}, although there are TDEs that exhibit hardness variability. For example, ASASSN-19dj shows similar "harder-when-fainter" behavior during the late time X-ray flare \citep{Hinkle2021}. The TDE candidate AT2018fyk \citep{Wevers2019,Wevers2021} also shows significant increase in its spectral hardness during the first two years after the flare. 

The black-body temperature $kT$ spans a range of $0.1 - 0.2$ keV. This is higher than the black-body temperature of X-ray bright TDEs such as ASASSN-14li \citep[e.g.,][]{Brown2017}, ASASSN-15oi \citep[e.g.,][]{Holoien2018} and ASASSN-19dj \citep{Hinkle2021}. The increase of $kT$ in the early epochs also differs from ASASSN-14li and ASASSN-19dj, where $kT$ declines as the lightcurve decays and ASASSN-15oi where $kT$ varies little with time. The black-body radius $R_{\rm bb} \sim 10^{11}$ cm is smaller than the Schwarzschild radius. This unphysically small $R_{\rm bb}$ is similar to the ANT ASASSN-18jd \citep{Neustadt2020}, while it differs from the TDEs ASASSN-14li and ASASSN-15oi where $R_{\rm bb}$ is larger than the Schwarzschild radius. This is more consistent with the AGN scenario where the black-body component is not associated with an accretion disk. The significant drop of $R_{\rm bb}$ is also inconsistent with these TDEs whose X-ray emission region sizes show little short time scale variability. \citet{Mummery2021b_Rp} presented a new time-dependent disk model that accounted for the disk temperature profile and disk opacity effects \citep[e.g.,][]{Done2012}. We obtained a disk radius $\sim 10^{12.5}$ cm using their model. However, the Schwarzschild radius ranges between $10^{11.9} - 10^{13.2}$ cm due to the large uncertainty of the black hole mass, so it is still uncertain whether this disk radius is larger than the Schwarzschild radius.

\citet{Frederick2021} presented infrared data and optical spectra of AT2019pev. The Wide-field Infrared Survey Explorer (WISE) color $W1 - W2 = 0.45$ supports the TDE scenario rather than AGNs \citep[e.g.,][]{Stern2012}. The H$\beta$ line width ${\rm FWHM}_{{\rm H}\beta} = 878$ km s$^{-1}$ disfavors TDEs, which generally have Balmer lines with ${\rm FWHM} \gtrsim 10^4$ km s$^{-1}$ \citep[e.g.,][]{Arcavi2014}. Nevertheless, the lack of the Fe II emission complex and the presence of strong He II line and Bowen fluorescence features are consistent with the TDE scenario \citep[e.g.,][]{vanVelzen2021}, although the ANT AT2019bgt also shows Bowen fluorescence features that were thought to be related to AGNs \citep{Trakhtenbrot2019}. The Eddington ratio of AT2019pev ranges between $0.066 - 1.5$ due to the order-of-magnitude difference between the two black hole mass estimates, so it is unclear whether AT2019pev is a super-Eddington accreting source. Both black hole mass estimates are smaller than the Hills mass $\sim 10^8 M_\odot$ \citep{Hills1975} for tidal disrupting a solar-type star, which is compatible with both the TDE and the AGN scenario.

\subsection{The AGN Scenario} \label{subsec:discussion_agn}
The existence of the ROSAT detection of AT2019pev supports the AGN scenario. AGN X-ray variability can be characterized by ``red noise'' with the power spectrum flattening toward low frequencies \citep[e.g.,][]{Lawrence1987,Uttley2002}. \citet{Auchettl2018} found that AGNs show large variations in the temporal power-law index $t^{-n}$, which is consistent with the stochastic nature of the AGN variability. The evolution of the temporal power-law index $n$ of AT2019pev is more consistent with AGNs than TDEs, although it does not show such a large variation as those in the \citet{Auchettl2018} sample except for the quick drop in the early epochs. The re-brightening in the UV/optical bands of AT2019pev is natural for AGNs with stochastic variability, although the amplitude is unusual \citep[e.g.,][]{MacLeod2010}.

A potential explanation is a flare created by disk instabilities \citep[e.g.,][]{Lightman1974}. In its quiescent state, the inner disk is truncated at some radius and the central region is either empty or filled with gas that has no significant X-ray emission. The inner disk is slowly refilled by the outer disk. When the radiation pressure exceeds the gas pressure, the inner disk generates a heating wave with a luminous flare. The inner disk is accreted faster than it can be re-filled by accretion from the outer disk, so the inner disk empties, fades and then starts a new cycle. This model has been used to interpret the flares in AGNs such as NGC 3599 \citep{Saxton2015} and IC 3599 \citep{Brandt1995,Grupe1995,Grupe2015}. IC 3599 shows repeated flares and a ``harder-when-fainter'' behavior, similar to that seen in AT2019pev. However, the second flare of IC 3599 occurs about 20 years after the first flare, while the second peak of AT2019pev is only about 200 days after the first peak. For AT2019pev, a luminous second peak is only seen in optical, while the re-brightening of X-ray emission is very modest.

The X-ray spectra of AGNs generally have relatively hard power-law indices with $\Gamma \sim 2$ \citep[e.g.,][]{Nandra1994}, so the early epoch spectra of AT2019pev with $\Gamma \sim 3 - 4$ are much softer than is typical for AGNs. However, NLSy1 galaxies can exhibit X-ray spectra with a power-law index of $\Gamma \sim 3$ \citep[e.g.,][]{Frederick2019}. The spectra of AT2019pev become harder as it fades after the peak. This "harder-when-fainter" behavior is extensively observed in luminous AGNs with a change of the Eddington ratio \citep[e.g.,][]{Grupe2004a,Grupe2004b,Shemmer2008,Grupe2010,Auchettl2018}. A potential explanation for this behavior is that Compton cooling of the X-ray corona becomes less efficient as the accretion rate of the disk decreases, leading to the production of fewer soft X-ray photons. The black-body temperature $kT \sim 0.1 - 0.2$ keV is consistent with non-blazar AGNs \citep[e.g.,][]{Ricci2017}. The strong He II line in the UV/optical spectra is atypical of AGNs, but there are exceptions such as the CL AGN AT2018dyk \citep{Frederick2019}. The range of the optical-to-X-ray slope $\alpha_{\rm OX} \sim 1.1 - 1.3$ is consistent with NLSy1s \citep[e.g.,][]{Gallo2006}.

\subsection{A Comparison with ANTs} \label{subsec:discussion_ant}
AT2019pev shows Bowen fluorescence features similar to the new class of transients such as AT2019bgt \citep{Trakhtenbrot2019}. However, both the UV/optical and X-ray lightcurves of AT2019pev drop steeply after the flare, in contrast to the slow evolution of the AT2019bgt lightcurves. AT2019pev also does not have a strong UV excess ($\alpha_{\rm OX} \sim 1.9$) like AT2019bgt. The X-ray spectra of AT2019bgt is described by a power-law with $\Gamma \sim 1.9$, which differs from the initially very steep X-ray spectra of AT2019pev. 

Compared to the ANT ASASSN-18jd, whose optical lightcurve decays by $\sim 0.5$ dex in a year after the peak, the lightcurve of AT2019pev is much steeper. The early spectra of ASASSN-18jd show strong He II and N III lines, but \citet{Neustadt2020} disfavor Bowen fluorescence as the origin of the N III lines due to the inconsistency of line ratios like O III $\lambda$3133 to O III $\lambda$3444 with this hypothesis. The X-ray lightcurves of ASASSN-18jd exhibit wide variations as it evolves, with a sharp flare at $\sim 140$ days after the optical peak, which is unlike the lightcurve of AT2019pev. The X-ray spectra of ASASSN-18jd flatten during the flare, similar to the "harder-when-fainter" behavior of AT2019pev. However, the softening of ASASSN-18jd is due to the change of the black-body component while the power-law is roughly constant, in contrast to the flattening power-law component of AT2019pev. 

The luminosity of the ANT ASASSN-20hx drops by $\sim 0.3$ dex in about 250 days after the peak, which is also a much slower decline than AT2019pev. The featureless spectra of ASASSN-20hx significantly differ from AT2019pev. The power-law index $\Gamma$ of the X-ray spectra of ASASSN-20hx evolves in a similar range to AT2019pev and it shows a similar, although weaker, "harder-when-fainter" trend. However, the X-ray lightcurve of ASASSN-20hx has more short time scale variability than AT2019pev.

\subsection{Uncommon Features} \label{subsec:discussion_uncommon}
The three pre-peak epochs of AT2019pev are very different from the post-peak epochs. They occupy a different area in the correlation between $\Gamma$ or HR and $L_X$ (see Figure \ref{fig:pvar}), because the spectra harden rather than soften during the rise to the peak. Figure \ref{fig:ufspec} shows that the luminosity increase before peak is mainly due to the increasing hard emission. The earliest \textit{Swift} epoch can be well-described by a black-body without the power-law component, while the latter two NICER epochs require a power-law to fit the hard tail. This may indicate the appearance of a discrete power-law component. The black-body component becomes hotter with roughly constant radius in these epochs, which could also contribute to the hardening. This "harder-when-brighter" behavior is uncommon for both TDEs and luminous AGNs. Variable obscuration is a potential reason for changes in the spectra shape, but our spectral models show that the absorption column density is consistent with the Galactic value without evolution over time, consistent with the lack of changes in absorption seen for TDEs \citep[e.g.,][]{Auchettl2017}. Both the before and after peak evolution of AT2019pev are likely to be intrinsic of the source.

Another explanation of the different evolution before and after the peak is a transition in the accretion state. X-ray sources associated with stellar-mass black holes show transitions among the soft, intermediate and hard accretion states \citep[e.g.,][]{Remillard2006}, and the spectral hardness shows different correlations with the luminosity (accretion rate) in different accretion states \citep[e.g.,][]{Wu2008}. SMBHs also show evidence for accretion state transitions. For example, \citet{Constantin2009} found an anti-correlation between the power-law index $\Gamma$ of the X-ray spectra and the Eddington ratio $L/L_{\rm Edd}$ (i.e., the "harder-when-brighter" behavior) for a sample of low-luminosity AGNs, while luminous AGNs show a positive $\Gamma - L/L_{\rm Edd}$ correlation with a turning point at $L/L_{\rm Edd} \approx 0.01$. 

The change of the correlation slope could be due to the transition from an advection-dominated accretion flow \citep[ADAF, e.g.,][]{Narayan1994} to a standard geometrically thin, optically thick disk \citep[e.g.,][]{Shakura1973} as the accretion rate increases. The dominant cooling mechanism for an ADAF is Comptonization of photons from the electron synchrotron radiation. The optical depth of the ADAF increases with the increase of the accretion rates, which in turn increases the Compton y-factor that characterizes the energy gain of the photons during the Compton scattering and makes the X-ray emission harder \citep[i.e., the "harder-when-brighter" behavior, e.g.,][]{Esin1997}. A standard thin disk starts to dominate as the accretion rate continues to increase, and the X-ray emission then shows the conventional "harder-when-fainter" behavior for luminous AGNs due to the Compton cooling mechanism discussed in Section \ref{subsec:discussion_agn}. 

This scenario can explain the different behaviors before and after peak. However, the transition from an ADAF to a standard thin disk requires a significant change in the accretion rate and luminosity, and we did not see such a change in AT2019pev. Another difference is that the ADAF model is generally applied to the hard states of XRBs, while for AT2019pev the spectra are quite soft before the peak. The host galaxy of AT2019pev was classified as a NLSy1 galaxy \citep{Frederick2021}, consistent with its relatively soft X-ray spectra. The accretion state transition scenario may imply a "turn-on" event from low accretion rate systems, such as low ionization nuclear emission regions (LINERs), to NLSy1 galaxies. For the LINER to NLSy1 "turn-on" event ZTF18aajupnt/AT2018dyk, \citet{Frederick2019} derived a power-law index $\Gamma \sim 3$ for the co-added X-ray spectra, which is consistent with the range of power-law index of AT2019pev. However, the X-ray flare of AT2018dyk is dominated by the soft bands, in contrast to the hard flare of AT2019pev. 
 
Transitions in accretion states have also been observed in TDEs. For example, the X-ray spectra of ASASSN-19bt soften as the lightcurve decays (i.e., the "harder-when-brighter" behavior), which may indicate a transition from hard to soft accretion state \citep{Holoien2019_19bt}. However, the "harder-when-brighter" behavior of ASASSN-19bt appears after the lightcurve peak, in contrast to AT2019pev, which shows this behavior before the peak. \citet{Wevers2021} found a transition from a soft state to a hard state, and then from the hard state to a quiescent state for the TDE candidate AT2018fyk. They attributed the former transition to forming an X-ray corona that produces stronger power-law components in the spectra. ASASSN-19dj has an X-ray flare $\sim 8$ months after discovery followed by a radio flare, which could indicate an accretion state transition \citep{Hinkle2021,Sfaradi2022}. However, the potential state transitions of both AT2018fyk and ASASSN-19dj happen a few months after the optical peak, while the transition of AT2019pev happens around the peak. It is unclear whether changes in the accretion states can be used to distinguish the AGN or a TDE in an AGN scenario.

%Summary
\section{Summary and Conclusion} \label{sec:summary}
We present extensive X-ray follow-up observations of the nuclear transient AT2019pev by Swift XRT, Chandra HRC/LETG, Chandra ACIS and NICER. The X-ray luminosity increases by a factor of five in $\sim 5$ days from the first \textit{Swift} epoch to the peak. It decays by a factor of $\sim 10$ with steeper slopes in early epochs and then flattens with a weak re-brightening trend after $t \sim 105$ days. The X-ray peak is very narrow compared to the UV/optical. The FWHM of the peak in the optical lightcurve is about 35 days, while for the X-rays it is only 7 days. The lightcurve between $t=27.6 - 105$ days can be fit by a power-law $L \propto t^{-n}$ with $n=1.67$. We fit the X-ray spectra with a power-law + black-body model and obtain a power-law index $\Gamma \sim 2 - 4$, a black-body temperature $kT \sim 0.1 - 0.175$ keV and a black-body radius $R_{\rm bb} \sim 10^{11}$ cm. The X-ray spectra show "harder-when-brighter" behavior before the lightcurve peak and "harder-when-fainter" behavior after the peak. The Gaia optical lightcurve extends beyond the seasonal limit of the ground-based observations used by \citet{Frederick2021} and is still rising toward an equally bright or brighter peak 223 days after the optical discovery and has then faded when the source is observable again. 
 
Combining the X-ray and multi-wavelength properties, we conclude that AT2019pev more closely resembles an AGN. While the spectra of AT2019pev show Bowen fluorescence features similar to the new class of transients by \citet{Trakhtenbrot2019}, other properties distinguish AT2019pev from this class. AT2019pev is also distinguished from other ANTs such as ASASSN-18jd and ASASSN-20hx. The unusual evolution of the X-ray spectra may indicate a transition of accretion states for an individual SMBH, which is not commonly observed.

\begin{table*}
\small
\renewcommand{\arraystretch}{1.35}
\begin{tabular}{llllllll}
\hline
Epoch Name & Instrument & MJD$-$MJD$_0$ & $\Gamma$ & $kT$ & reduced $\chi^2$ & $L_X$ & HR \\
 &  & (${\rm MJD}_0 = 58727.5$) &  & (keV) & & ($10^{44}$ erg/s) &  \\
\hline
Sw00011566001 & Swift & 22.7 & $3.81^{+1.72}_{-0.34}$ & $0.101^{+0.021}_{-0.011}$ & 0.95 & $0.76^{+0.17}_{-0.05}$ & $-0.98^{+0.02}_{-0.01}$ \\
NI2200860101 & NICER & 24.2 & $3.61^{+0.08}_{-0.06}$ & $0.128^{+0.003}_{-0.002}$ & 1.08 & $2.27^{+0.02}_{-0.02}$ & $-0.94^{+0.01}_{-0.01}$ \\
NI2200860102 & NICER & 24.5 & $3.57^{+0.05}_{-0.06}$ & $0.130^{+0.003}_{-0.003}$ & 1.01 & $2.99^{+0.02}_{-0.02}$ & $-0.92^{+0.01}_{-0.01}$ \\
Sw00011566004 & Swift & 27.6 & $3.13^{+0.13}_{-0.14}$ & $0.136^{+0.020}_{-0.010}$ & 1.04 & $3.04^{+0.07}_{-0.07}$ & $-0.80^{+0.02}_{-0.02}$ \\
NI2200860105 & NICER & 27.6 & $3.25^{+0.03}_{-0.03}$ & $0.136^{+0.003}_{-0.003}$ & 1.00 & $3.60^{+0.02}_{-0.02}$ & $-0.84^{+0.01}_{-0.01}$ \\
NI2200860106 & NICER & 28.6 & $3.10^{+0.03}_{-0.04}$ & $0.151^{+0.005}_{-0.005}$ & 1.01 & $3.05^{+0.02}_{-0.02}$ & $-0.78^{+0.01}_{-0.01}$ \\
Sw00011566005 & Swift & 28.6 & $2.85^{+0.19}_{-0.14}$ & $0.134^{+0.022}_{-0.023}$ & 1.07 & $2.39^{+0.09}_{-0.08}$ & $-0.71^{+0.03}_{-0.03}$ \\
NI2200860107 & NICER & 29.6 & $3.08^{+0.04}_{-0.05}$ & $0.162^{+0.012}_{-0.009}$ & 1.09 & $2.44^{+0.03}_{-0.04}$ & $-0.77^{+0.01}_{-0.01}$ \\
NI2200860108 & NICER & 30.5 & $3.00^{+0.06}_{-0.06}$ & $0.153^{+0.010}_{-0.010}$ & 0.99 & $1.86^{+0.02}_{-0.02}$ & $-0.76^{+0.02}_{-0.02}$ \\
NI6-7 & NICER & 32.2 & $2.88^{+0.07}_{-0.07}$ & $0.162^{+0.009}_{-0.008}$ & 1.18 & $1.59^{+0.02}_{-0.03}$ & $-0.74^{+0.02}_{-0.02}$ \\
NI2200860111 & NICER & 33.5 & $2.79^{+0.07}_{-0.06}$ & $0.163^{+0.009}_{-0.009}$ & 0.84 & $1.58^{+0.03}_{-0.03}$ & $-0.71^{+0.02}_{-0.02}$ \\
Sw08-02 & Swift & 35.1 & $2.73^{+0.12}_{-0.28}$ & $0.141^{+0.019}_{-0.020}$ & 0.89 & $1.13^{+0.03}_{-0.07}$ & $-0.64^{+0.05}_{-0.03}$ \\
NI9-10 & NICER & 35.5 & $2.73^{+0.07}_{-0.08}$ & $0.167^{+0.013}_{-0.011}$ & 0.86 & $1.50^{+0.03}_{-0.02}$ & $-0.67^{+0.02}_{-0.03}$ \\
NI11-13 & NICER & 38.8 & $2.83^{+0.08}_{-0.11}$ & $0.162^{+0.009}_{-0.010}$ & 0.96 & $1.27^{+0.03}_{-0.03}$ & $-0.73^{+0.03}_{-0.03}$ \\
AC21390 & Chandra ACIS & 40.6 & $1.88^{+0.09}_{-0.09}$ & $0.133^{+0.004}_{-0.006}$ & 1.21 & $0.98^{+0.05}_{-0.03}$ & $-0.55^{+0.03}_{-0.02}$ \\
SeqN703702 & Chandra grating & 42.5 & $3.03^{+0.20}_{-0.18}$ & $0.162^{+0.044}_{-0.025}$ & 0.81 & $1.96^{+0.08}_{-0.08}$ & $-0.77^{+0.06}_{-0.05}$ \\
Sw03-05 & Swift & 44.1 & $2.52^{+0.31}_{-0.16}$ & $0.124^{+0.015}_{-0.015}$ & 0.86 & $1.32^{+0.04}_{-0.08}$ & $-0.65^{+0.03}_{-0.05}$ \\
NI14-15 & NICER & 50.1 & $2.83^{+0.06}_{-0.08}$ & $0.154^{+0.007}_{-0.007}$ & 1.04 & $1.19^{+0.02}_{-0.01}$ & $-0.72^{+0.02}_{-0.02}$ \\
NI16-17 & NICER & 54.3 & $2.42^{+0.20}_{-0.16}$ & $0.134^{+0.016}_{-0.008}$ & 1.07 & $1.00^{+0.04}_{-0.05}$ & $-0.57^{+0.06}_{-0.05}$ \\
Sw06-09 & Swift & 54.4 & $2.24^{+0.35}_{-0.24}$ & $0.149^{+0.017}_{-0.020}$ & 0.85 & $0.79^{+0.05}_{-0.05}$ & $-0.56^{+0.06}_{-0.07}$ \\
SeqN703703 & Chandra grating & 55.5 & $2.96^{+0.21}_{-0.30}$ & $0.146^{+0.020}_{-0.019}$ & 0.14 & $1.00^{+0.08}_{-0.08}$ & $-0.76^{+0.05}_{-0.04}$ \\
NI18-20 & NICER & 58.8 & $2.55^{+0.10}_{-0.12}$ & $0.146^{+0.007}_{-0.005}$ & 0.89 & $0.85^{+0.02}_{-0.02}$ & $-0.65^{+0.03}_{-0.03}$ \\
NI2200860125 & NICER & 62.5 & $2.80^{+0.10}_{-0.12}$ & $0.149^{+0.011}_{-0.010}$ & 1.08 & $0.90^{+0.03}_{-0.03}$ & $-0.71^{+0.03}_{-0.03}$ \\
NI22-23 & NICER & 64.7 & $2.66^{+0.13}_{-0.22}$ & $0.140^{+0.007}_{-0.006}$ & 0.90 & $0.84^{+0.03}_{-0.02}$ & $-0.73^{+0.04}_{-0.03}$ \\
Sw10-13 & Swift & 65.0 & $2.44^{+0.37}_{-0.23}$ & $0.124^{+0.012}_{-0.012}$ & 0.86 & $0.92^{+0.05}_{-0.05}$ & $-0.62^{+0.06}_{-0.06}$ \\
NI24-25 & NICER & 70.2 & $2.76^{+0.17}_{-0.14}$ & $0.142^{+0.026}_{-0.014}$ & 1.12 & $0.60^{+0.02}_{-0.02}$ & $-0.69^{+0.04}_{-0.04}$ \\
SeqN703704 & Chandra grating & 70.5 & $1.39^{+1.45}_{-1.37}$ & $0.138^{+0.013}_{-0.010}$ & 0.71 & $1.03^{+0.29}_{-0.16}$ & $-0.51^{+0.50}_{-0.24}$ \\
AC21391 & Chandra ACIS & 72.3 & $2.11^{+0.11}_{-0.15}$ & $0.141^{+0.011}_{-0.009}$ & 0.78 & $0.57^{+0.03}_{-0.04}$ & $-0.57^{+0.03}_{-0.03}$ \\
Sw14-19 & Swift & 78.7 & $2.36^{+0.36}_{-0.26}$ & $0.133^{+0.021}_{-0.017}$ & 0.83 & $0.58^{+0.03}_{-0.04}$ & $-0.60^{+0.06}_{-0.06}$ \\
SeqN703705 & Chandra grating & 81.5 & $2.97^{+0.24}_{-0.84}$ & $0.143^{+0.023}_{-0.017}$ & 0.36 & $0.79^{+0.07}_{-0.07}$ & $-0.75^{+0.16}_{-0.06}$ \\
NI26-28 & NICER & 87.2 & $2.56^{+0.15}_{-0.19}$ & $0.134^{+0.013}_{-0.007}$ & 1.24 & $0.70^{+0.02}_{-0.03}$ & $-0.62^{+0.06}_{-0.04}$ \\
NI2200860134 & NICER & 98.3 & $2.69^{+0.27}_{-0.35}$ & $0.131^{+0.020}_{-0.010}$ & 0.94 & $0.43^{+0.03}_{-0.02}$ & $-0.71^{+0.09}_{-0.06}$ \\
Sw21-29 & Swift & 104.5 & $2.52^{+0.25}_{-0.22}$ & $0.136^{+0.019}_{-0.012}$ & 0.80 & $0.32^{+0.02}_{-0.02}$ & $-0.59^{+0.05}_{-0.05}$ \\
NI30-32 & NICER & 110.9 & $2.53^{+0.18}_{-0.22}$ & $0.136^{+0.013}_{-0.008}$ & 0.89 & $0.43^{+0.02}_{-0.02}$ & $-0.62^{+0.05}_{-0.05}$ \\
Sw31-39 & Swift & 138.5 & $2.28^{+0.23}_{-0.18}$ & $0.154^{+0.023}_{-0.021}$ & 0.91 & $0.38^{+0.02}_{-0.02}$ & $-0.52^{+0.06}_{-0.06}$ \\
AC21392 & Chandra ACIS & 139.3 & $1.93^{+0.14}_{-0.11}$ & $0.156^{+0.013}_{-0.012}$ & 1.19 & $0.45^{+0.03}_{-0.03}$ & $-0.49^{+0.02}_{-0.03}$ \\
NI2200860139 & NICER & 160.9 & $2.49^{+0.17}_{-0.18}$ & $0.142^{+0.013}_{-0.010}$ & 0.92 & $0.55^{+0.02}_{-0.03}$ & $-0.61^{+0.05}_{-0.05}$ \\
Sw40-46 & Swift & 174.2 & $2.33^{+0.29}_{-0.26}$ & $0.141^{+0.049}_{-0.020}$ & 1.01 & $0.39^{+0.02}_{-0.02}$ & $-0.52^{+0.06}_{-0.06}$ \\
NI34-39 & NICER & 181.5 & $2.73^{+0.12}_{-0.12}$ & $0.159^{+0.027}_{-0.016}$ & 0.71 & $0.44^{+0.01}_{-0.02}$ & $-0.67^{+0.05}_{-0.04}$ \\
NI3200860104 & NICER & 195.7 & $2.80^{+0.50}_{-0.58}$ & $0.153^{+0.041}_{-0.024}$ & 0.88 & $0.29^{+0.03}_{-0.02}$ & $-0.71^{+0.13}_{-0.15}$ \\
\hline
\end{tabular}
\caption{Epoch properties and model parameters. Columns (1) and (2) give the epoch name and instrument. Column (3) gives the time since discovery in units of days. Columns (4) through (8) give the power-law index, the black-body temperature, the reduced $\chi^2$, the X-ray luminosity at 0.3$-$5 keV and the hardness ratio. The uncertainties correspond to 90\% confidence level.}
\label{tab:param}
\end{table*}

\section*{Acknowledgements}
%
%The Acknowledgements section is not numbered. Here you can thank helpful
%colleagues, acknowledge funding agencies, telescopes and facilities used etc.
CSK is supported by NSF grants AST-1908570 and AST-1814440. CSK, SM and ZY are supported by Chandra grant GO9-20084X. Parts of this research were supported by the Australian Research Council Centre of Excellence for All Sky Astrophysics in 3 Dimensions (ASTRO 3D), through project number CE170100013. Support for T.W.-S.H. was provided by NASA through the NASA Hubble Fellowship grant HST-HF2-51458.001-A awarded by the Space Telescope Science Institute (STScI), which is operated by the Association of Universities for Research in Astronomy, Inc., for NASA, under contract NAS5-26555.

We thank the PI, the Observation Duty Scientists, and the science planners of \textit{Swift} and NICER for promptly approving and executing our observations. We acknowledge the use of public data from the Swift data archive. Support for this work was provided by the National Aeronautics and Space Administration through Chandra Award Number GO9-20084X issued by the Chandra X-ray Center, which is operated by the Smithsonian Astrophysical Observatory for and on behalf of the National Aeronautics Space Administration under contract NAS8-03060. The scientific results reported in this article are based in part on observations made by the Chandra X-ray Observatory. This research has made use of data obtained from the Chandra Data Archive and the Chandra Source Catalog, and software provided by the Chandra X-ray Center (CXC) in the application packages CIAO and Sherpa. This research has made use of data and/or software provided by the High Energy Astrophysics Science Archive Research Center (HEASARC), which is a service of the Astrophysics Science Division at NASA/GSFC. 

We acknowledge ESA Gaia, Gaia Data Processing and Analysis Consortium (DPAC) and the Photometric Science Alerts Team (http://gsaweb.ast.cam.ac.uk/alerts). Gaia data are being processed by the DPAC. Funding for the DPAC is provided by national institutions, in particular the institutions participating in the Gaia MultiLateral Agreement (MLA). The Gaia mission website is https://www.cosmos.esa.int/gaia. The Gaia archive website is https://archives.esac.esa.int/gaia.

%Data Availability 
\section*{Data Availability}
All X-ray data used in this paper are available on HEASARC (https://heasarc.gsfc.nasa.gov/docs/archive.html).

%%%%%%%%%%%%%%%%%%%%%%%%%%%%%%%%%%%%%%%%%%%%%%%%%%



\bibliographystyle{mnras}
\bibliography{ref} % if your bibtex file is called example.bib

%\begin{thebibliography}{99}
%\end{thebibliography}

%%%%%%%%%%%%%%%%%%%%%%%%%%%%%%%%%%%%%%%%%%%%%%%%%%

%%%%%%%%%%%%%%%%% APPENDICES %%%%%%%%%%%%%%%%%%%%%

%\appendix



%%%%%%%%%%%%%%%%%%%%%%%%%%%%%%%%%%%%%%%%%%%%%%%%%%

% Don't change these lines
\bsp	% typesetting comment
\label{lastpage}
\end{document}